# Shear-band cavitation determines the shape of the stress-strain curve of metallic glasses

A. Das[1,2], C. Ott[1], D. Pechimuthu[3], R. Moosavi[3], M. Stoica[4], P.M. Derlet[5], R. Maaß[1,3*]

[1] Department of Materials Science and Engineering, University of Illinois at Urbana-Champaign, Urbana, IL 61801, USA

[2] Cornell High Energy Synchrotron Source, Cornell University, Ithaca, NY 14853, USA

[3] Federal Institute of Materials Research and Testing (BAM), Unter den Eichen 87, 12205 Berlin, Germany

[4] Laboratory of Metal Physics and Technology, Department of Materials, ETH Zurich, 8093 Zurich, Switzerland

[5] Condensed Matter Theory Group, Paul Scherrer Institute, Villigen, PSI, Switzerland

Metallic glasses are known to have a remarkably robust yield strength, admitting Weibull moduli as high as for crystalline engineering alloys. However, their post-yielding behavior is strongly varying, with large scatter in both flow stress levels and strains at failure. Using x-ray tomography we reveal for the first time how a strain-dependent internal evolution of shear-band cavities underlies this unpredictable post yielding response. We demonstrate how macroscopic strain-softening coincides with the first detection of internal shear-band cavitation. Cavity growth during plastic flow is found to follow a power-law, which yields a fractal dimension and a roughness exponent in excellent agreement with self-similar surface properties obtained after fracture. These findings demonstrate how internal micro-cracking coexists with shear-band plasticity along the plastic part of a stress-strain curve, rationalizing the large variability of plastic flow behavior seen for metallic glasses.

*Corresponding author. Email address: robert.maass@bam.de (R. Maaß)

1. Introduction

How the stress of a material changes with strain is a central concept in materials science. A large number of fundamental mechanical properties can be derived from a stress-strain curve, and if recorded while applying complex loading protocols with changing strain rates or temperatures, underlying deformation mechanisms and microstructural changes can be assessed in the deforming material. As such, recording stress-strain curves is elementary to all mechanical property determination and is a founding principle for mechanical metallurgy and structural applications.

It is customary to represent stress and strain with respect to either the original specimen geometry, yielding the engineering stress-strain curve, or to take into account the gradually changing load-bearing area and length that returns the well-known true stress-strain response. Whilst the latter case can be conveniently calculated for polycrystalline metals and alloys that distribute plasticity homogenously throughout the deforming volume, highly localized shear-deformation confined to nano-scale shear-bands in metals with a disordered atomic structure, i.e. metallic glasses (MGs), typically restricts flow curves to the engineering stress-strain type [1-3].

Relying on engineering stress-strain data is a reasonably good approximation for small plastic strains and obviously for the microplastic pre-yield regime [4], in which engineering and true quantities macroscopically do not differ. As such, engineering stress strain curves are used in analysing component reliability, where plastic deformation is typically not expected to occur. With increasing plastic strain, the deviations between engineering and true quantities diverge significantly. In the case of MGs, where shear bands mediate plastic strain accumulation, the shear-event magnitude from each active shear band can be measured and accounted for [5-7], thereby giving access to the geometrically corrected stress. In particular, the case of serrated inhomogeneous flow offers tracking each abrupt increase in strain through the intermittent activity of shear bands and an apparent true stress-strain curve can be constructed, which we later will refer to as the 'semi-true' stress-strain curve. Whilst possible, such detailed tracking of shear instabilities of MGs is rarely done and tends to only correct stress-strain values by negligible amounts due to small strains and limited geometric changes at failure.

In addition to geometrical considerations that may dictate differences between engineering and true stress-strain quantities, it is further interesting to note that a series of nominally identical MG-samples and testing protocols can result in a large variety of post yielding behaviour [8]. Figure 1 demonstrates this for 6 mechanical tests on a $Zr_{52.5}Cu_{17.9}Ni_{14.6}Al_{10}Ti_5$ (Vit105) bulk MG recorded with a screw-driven universal testing machine in compression at a strain rate of $10^{-3}\ s^{-1}$. Despite experimentally identically conditions, such variations are common in MGs and are much larger than for conventional crystalline metals but are rarely collectively represented in the literature. What is the fundamental origin of this variable post-yielding behaviour? A closer inspection reveals that under well-controlled testing conditions the fluctuations in yield stress of a ductile MG can be comparable to some crystalline engineering alloys with Weibull moduli as high as 75 [8]. This is remarkable on its own, demonstrating that despite the amorphous structure, MGs are not necessarily brittle in the sense of brittle oxide glasses or ceramics. Indeed, yielding is characterized by a robust elastic-to-plastic transition that can be captured with a thermally-activated plastic transition state theory [9, 10]. In contrast to well-reproducible yielding, a variety of post-yield behaviours has been reported, including strain softening [11], strain hardening [12-15], perfect plastic flow [16, 17], or mixtures of these evolutions in the same deformation curve [18, 19]. Furthermore, large differences in strain at failure can be observed (for example a factor of 2, as seen in Figure 1), which becomes even more pronounced when testing under misaligned or geometrically constrained conditions [8, 20]. In particular, the case of strain hardening has been subject of debate since it remains unclear how a material that lacks processes of structural defect interaction and defect entanglement can intrinsically harden [12, 13].

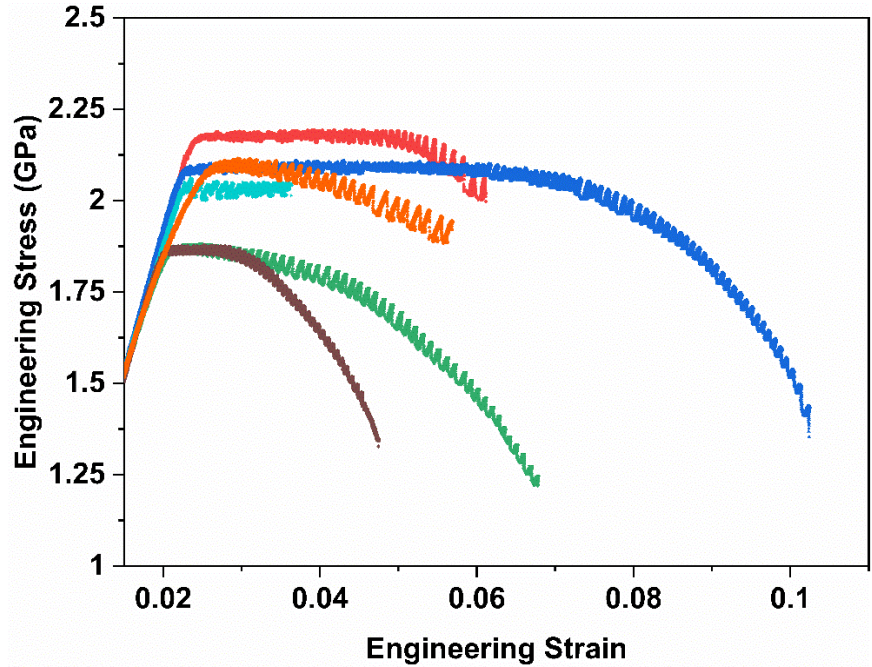

Figure 1: Stress-strain curves of a $Zr_{52.5}Cu_{17.9}Ni_{14.6}Al_{10}Ti_5$ MG obtained under experimentally identical conditions. This set exemplifies the typical variability in post yielding behavior. We note that the stress axis begins at 1 GPa and that the yield-stress fluctuations therefore are amplified.

One step forward in our understanding of what could cause the variability in flow behaviour was provided by recent x-ray tomography work [11, 21-23], highlighting for a series of deformed MG specimens the presence of internal voids, also referred to as shear-band cavities. These are located on the shear plane inside the deforming sample. The area fraction of such shear-band cavities can vary substantially amongst samples and can amount to values larger than 90% within intact (unbroken) compression specimens [11]. Being observed *after* mechanical testing, it is possible that such shear-band cavities are a result of a stress-relaxation process following plastic flow, which would decouple the observed voids from the actual plastic flow of the material. Currently available insights thus reflect the end of a structural damage process that presumably begins with sub-nanoscale void formation [24-26] within shear bands and ends with fracture. Whilst the formation and evolution of internal microcracks on the shear-band plane has been alluded to, factual evidence of shear-band cavitation (internal micro-cracking) is limited to large strains when samples have remained intact by fortuitously achieving some appropriate structural realization. It remains consequently entirely unclear how internal shear-band cavities evolve with strain and how a strain-dependent correction of this internal decohesion process would manifest itself in the shape of the plastic flow curve. In other words, revealing the detailed internal shear-band cavity structure as a function of strain, combined with correcting for the geometric changes due to localized shear, would for the first time give access to the true stress-strain curve of a MG and possibly shed light on the long-standing question why the post-yielding response varies as dramatically, as exemplified in Figure 1.

To this end, we pursue here a combined x-ray tomography (XRT) and in-situ acoustic emission (AE) study to reveal the detailed and strain-dependent formation and growth of shear-band cavities, with the final goal of calculating a true stress-strain curve of a MG. A series of step-wise loading segments and 50 XRT scans show how shear-band cavities form at various locations on the shear-band plane and at different plastic strains. These shear-band cavities grow approximately exponential with strain and exhibit fractal behaviour, the latter of which can be shown via a power-law scaling between the cavity-area and perimeter. Both the fractal dimension and the therefrom derived roughness exponent are in remarkable agreement with what is known from post-mortem fracture surface analysis. First macroscopic mechanical softening is observed when resolvable shear-band cavities are detected. Tracing their strain-dependent evolution allows calculating the first true stress-strain curve of a metallic glass. Even when fully corrected for geometry changes due to inhomogeneous deformation and for the internal reduction in load-bearing area, strain softening is seen, which therefore reflects the isolated intrinsic strain-dependent structural softening (damage or free volume accumulation) of the shear-band material. These results demonstrate that fracture and plastic shear deformation co-exist over a large part of a compressive stress-strain curve from a metallic glass and link the shape of the stress strain curve to the detailed evolution of shear-band cavities.

2. **Materials and methods**

Cylindrical MG rods of composition $Zr_{65}Cu_{25}Al_{10}$, and $2\ mm$ diameter were produced by suction casting into a Cu mould. Samples of $4.5\ mm$ height were cut from the rod and polished using a custom jig to obtain plane parallel geometry. A notch was created on the sample using electro discharge machining to ensure the formation of a single, macroscopic shear band that spans across the width of the sample (system spanning) and in which all macroscopic plastic strain is localized. Earlier work has shown that this yields identical results to single shear banding without notching [27]. In addition to the data in Fig. 1, one sample was tested in compression using a screw-driven uniaxial testing machine (UTM), coupled with in-situ acoustic emission (AE) testing. In total, 49 successive load-reload experiments were conducted. The AE setup comprises of a PICO sensor (operating frequency range: 250 – 700 $kHz$) and a preamplifier (gain: 60 $dB$), read-out by a data acquisition card with a sampling rate of 2 $MHz$. The AE data is time-synchronized to the load-displacement data from the mechanical test by feeding the load signal from the Instron as a parametric channel into the AE data acquisition card. Further details on the AE set-up can be found in Ref. [28]. Compression tests for this sample were performed at a displacement rate of 2.5 $\times 10^{-3}$ $mm/s$. Fiducial markers were used to ensure uniform sample alignment with respect to the compression anvils across multiple mechanical tests. Stress-strain curves are calculated in three different ways: 1) The measured load is converted to engineering stress ($\sigma_e$) by dividing it with the cross-sectional area of the undeformed sample ($\sigma_e = F/A_{undef}$). The sample displacement is corrected for UTM compliance and the engineering strain ($\varepsilon_e$) is calculated by dividing the sample displacement ($\Delta l$) with the length of the undeformed sample ($\varepsilon_e = \Delta l/l_0$). 2) A semi-true stress ($\sigma_{s-t}$) is calculated under the presumption that the plastic strain is accommodated in one system-spanning shear band. The semi-true cross-sectional area ($A_{s-t}$) as a function of strain is found by calculating the overlap between the elliptical shear-plane cross sections as the sample is progressively strained, allowing us to calculate semi-true stress ($\sigma_{s-t} = F/A_{s-t}$). The instantaneous length ($l$) of the sample is calculated by considering the plastic strain accumulated on the system spanning shear band and the true strain is calculated ($\varepsilon_t = \Delta l/l$). 3) A true stress is calculated using the true cross-sectional area by subtracting the cavity area ($A_{cav}$) from the semi-true cross-sectional area ($A_t = A_{s-t} - A_{cav}$), yielding the true stress $\sigma_t = F/A_t$. Each mechanical test was limited to a total strain equivalent of one or two stress-drop (serrations), with the exception of tests #17, #18, #20 and #21. The four exceptions were chosen so as to accelerate the development of shear-band cavities, which were very rare until that point in the deformation process. Between each mechanical test, XRT scans were performed using an Xradia MicroXCT-200. X-rays of a photon energy of 149 $keV$ were used to conduct tomography in absorption contrast mode. The sample and detector positions, as well as the objective magnification, are chosen to fit the entirety of the shear-plane area of interest and together correspond to a voxel size of 2.1 $\mu m$, which is ca. 0.6 $\mu m$ larger than in previous related synchrotron XRT work [11]. Every tomography scan comprises of 986 z-slices, each $984 \times 1005$ pixels in size, which were analyzed using the Dragonfly deep learning algorithm [29]. The deep learning algorithm was able to distinguish between the bulk of the MG, the air surrounding the sample and lastly, the shear-band cavities inside the MG. Location-specific filtering was used to remove voxels falsely identified as shear-band cavities. 3-D reconstruction of the scans confirmed that shear-band cavities are only present along the system-spanning shear band. A 2D projection looking top-down onto the shear plane is used to facilitate tracking of the cavities as a function of strain.

## 3. Results

### 3.1 Load-reload behaviour and strain-dependent shear-band cavity evolution

Figures 2a and 2b summarize the 49 consecutive loading segments that the here investigated sample was subjected to. Using a precisely defined sample positioning relative to the compression anvils, it is within experimental precision possible to reload the specimen after every XRT scan without notably changing the mechanical boundary conditions. As a result, it is possible to later combine all loading segments to one continuous serrated stress-strain curve. At a total true axial strain of ca. $\varepsilon_t = 0.021$ a deviation from the linear elastic regime is seen, which originates from the formation of a few non-system spanning shear bands. True strain refers here to the strain calculated based on the actual momentary length after each serration (Section 2), and will be used throughout the manuscript, even though the corresponding true stress-strain curve is first presented later in Figure 5. With continued loading, a system spanning shear band develops within loading segment #1 and the engineering stress-strain curves exhibits minimal strain-softening till a corresponding true strain of $\varepsilon_t = 0.068$ (loading segments #1-18) with a gradual decrease in the stress level. As will be shown later, this does not originate solely from the cross-sectional change due to the advancement of the upper sample half relative to the lower.

Already at small strains shear-band cavities of dimensions at the XRT resolution limit contribute to a reduction in loadbearing area. Since the shear-band cavity area-fraction tracked with XRT continued to be small and only grew slowly, loading segments #17, #18, #20 and #21 were used to accumulate larger amounts of strain per loading. In Figure 2a it can clearly be seen that loading segment #18 is the first to display a marked increase in the strain softening response that continues until loading segment #22. Further loading exhibits an overall linearly decreasing flow stress level in the engineering stress-strain representation. The last loading segment #49 reaches an engineering yield stress of 630 MPa, which corresponds to a reduction of 56% relative to the first elastic-to-plastic transition.

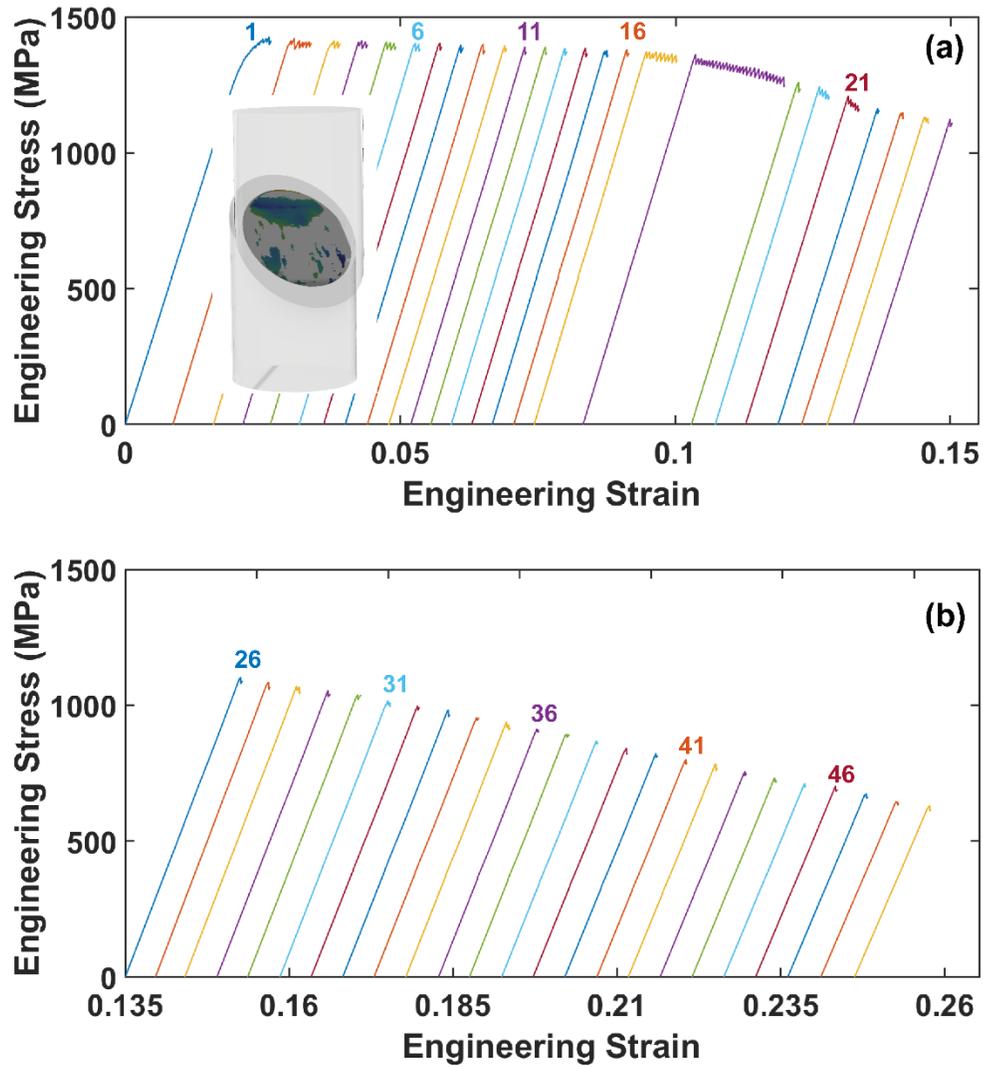

Figure 2: Engineering stress-strain curves obtained from 49 consecutive compression tests of one and the same sample. Typically, one to two stress drops are accumulated in each mechanical test and x-ray tomography scans are performed after each test. In-situ AE sensing was also used throughout all tests. (a) shows the first 25 tests and (b) shows the latter 24. Selected loading segments are labelled with their test number. The stress-strain curves of test #2-49 are shifted in strain for better visibility. An inset in (a) visualized schematically the shear-band plane after all loading segments, where colored areas correspond to shear-band cavities.

Given 49 mechanical tests, a total of 50 tomography scans were conducted, allowing to track the strain-dependence of the internal shear-band cavity evolution, which furthermore gives access to the true load bearing area after each reloading. Following segmentation and projection of the shear-band plane along its normal vector onto a 2D plane, the detailed shear-band cavity structure within the sample is tracked, as exemplified in Figure 3. Initial internal microcracking emerges on the left-hand side of the projected view shown in Figure 3a. First resolvable shear-band cavity formation is detected at a plastic true strain $\varepsilon_{pl} = 0.024$ but remains weak until $\varepsilon_{pl} = 0.03$, where we define $\varepsilon_{pl} = \varepsilon_t - 0.021$. A large increase in shear-band cavity area fraction occurs at a plastic strain of $\varepsilon_{pl} = 0.048$ (Figure 3a). This coincides with the marked mechanical softening seen in Figure 2a throughout loading segments #18-21. As the plastic strain increases, four different features can be distinguished from the full series of XRT scans: (i) shear-band cavities grow in area, (ii) new shear-band cavities appear, (iii) smaller shear-band cavities merge, and (iv) shear-band cavities may either remain internal or connect to the external surface, the morphology of which can be derived from the corresponding XRT images in Figures 3a-3f. To quantitatively study the strain-dependent shear-band cavity evolution, we track 5 individual ones (marked in 3d) and extract the shear-band cavity area, boundary length, volume, and average thickness. In the following section, we discuss the deformation induced changes of the first two parameters.

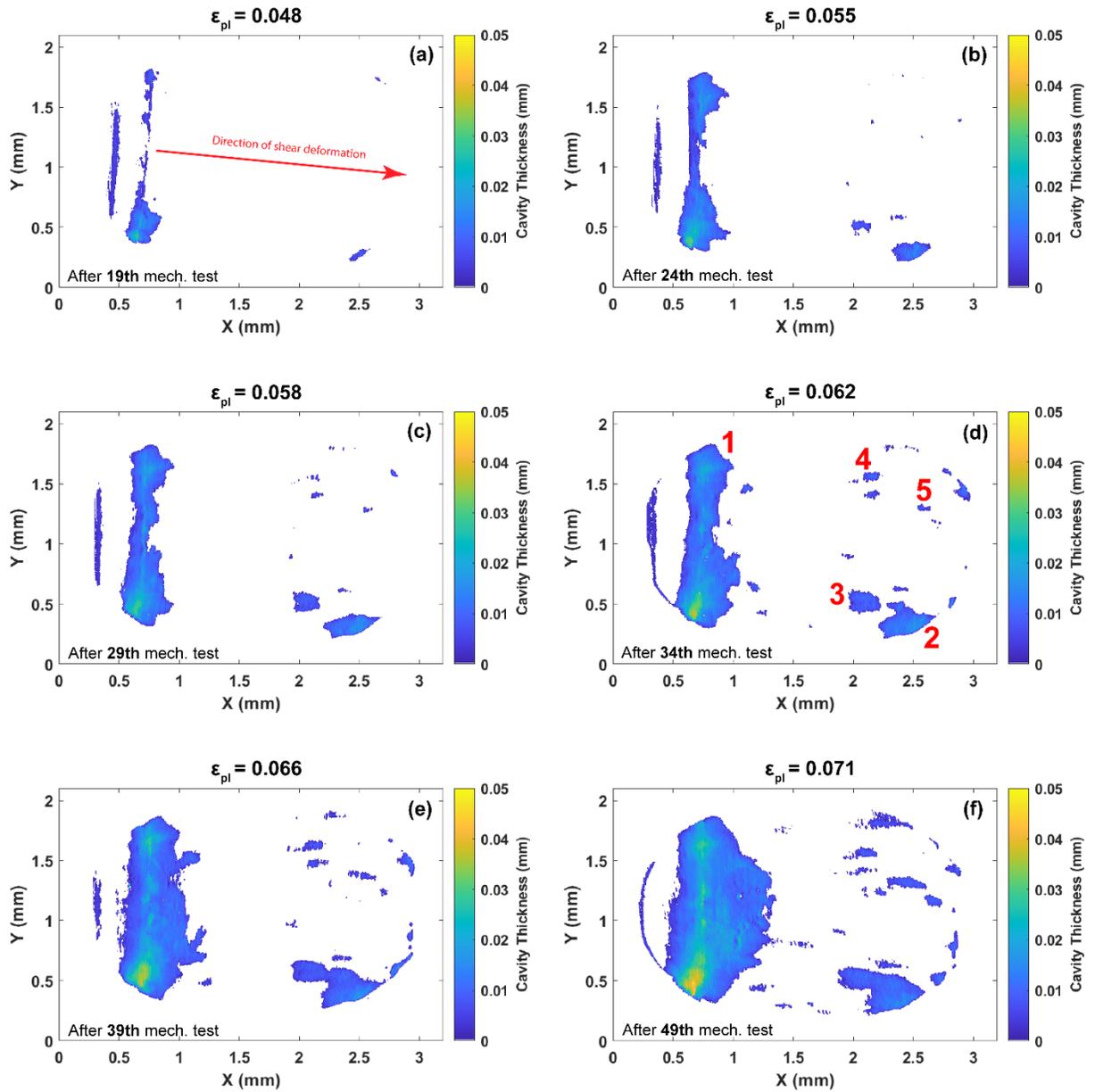

Figure 3: Examples of shear-band cavity morphologies as a function of plastic strain. The color code indicates the cavity thickness (height) and the shear direction is indicated in (a). The cavities marked in (d) are quantitatively analyzed as a function of strain (Figure 4).

Figure 4 summarizes the evolution of area and boundary length of the five shear-band cavities indicated in Figure 3d as a function of strain. Both quantities are displayed in Figures 4a and 4b and evidence a pronounced growth as a function of plastic strain. Fitting using various functional forms yields only a weak favour for an exponential growth even though the data appears as approximately linear. As can be seen in Figure 4a, the selected shear-band cavities form (within the resolution of the experiment) at different plastic strains. Specifically, the emergence of shear-band cavities 2 and 3, and 4 and 5 is separated by the accumulation of ~0.04% and ~0.08% plastic strain (1 or 2 stress drops), respectively. It is further seen that new shear-band cavities, as exemplified here with cavity nr. 2 and 3 for $\varepsilon_{pl} < 0.064$, may initially be isolated and can subsequently merge with nearby existing ones. This can also be deduced from Figures 3d,e,f. These observations show that local decohesion on the shear-band plane can occur at a variety of locations across the shear plane and that their growth and merging eventually leads to the large shear-band cavities observed prior to fracture [11, 23]. In other words, shear-band decohesion/cracking is not a propagating damage front that initiates at the sample surface and moves across the sample. Instead, it is controlled by a number of decohesion processes that are distributed across independent locations on the shear plane. Tracking individual shear-band cavities also reveals that growth in the direction of shear (indicated in Figure 3a) is more pronounced than orthogonal to the shear direction.

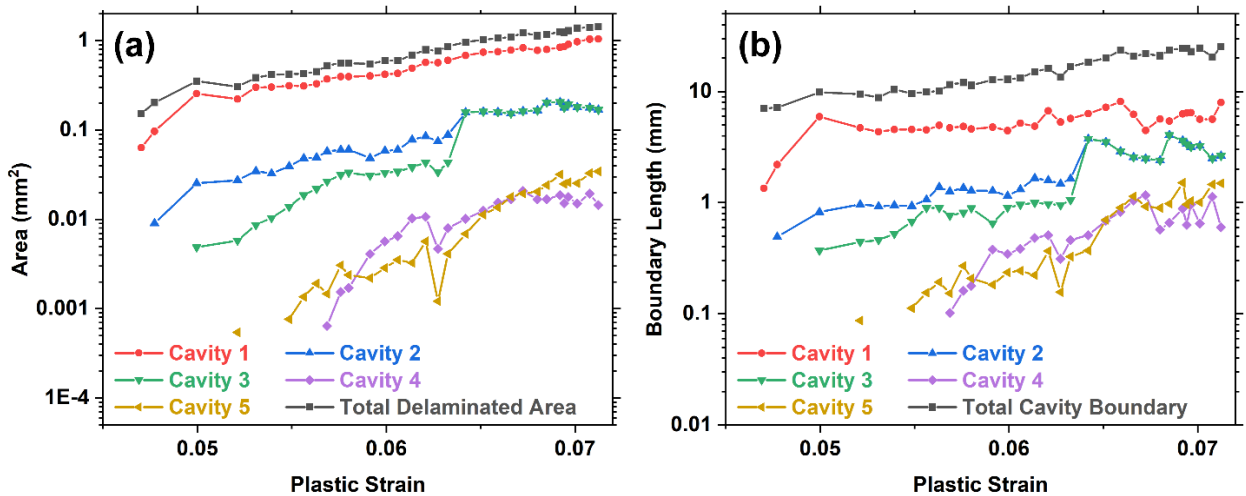

Figure 4: Shear-band cavity area (a) and boundary length (b) as a function of plastic strain for the five selected shear-and cavities highlighted in Figure 3d.

Another common feature of both the shear-band cavity area and boundary length is the presence of transients, where both abrupt increasing and decreasing trends are observed. This non-monotonic growth occurs multiple times as a function of plastic strain, but with different magnitudes. Two possible origins to the temporarily decreasing trends can be defined after a careful examination of the XRT data. The first one is based on the Dragonfly image segmentation algorithm, which determines if a particular voxel of the XRT data represents the bulk MG, a shear-band cavity, or the surrounding air. Under conditions when the algorithm sensitivity is conservatively chosen, it may occur that the edge regions of the bulk material and the shear-band cavity are classified as the adjacent bulk material, leading to an apparent temporary size reduction of a given shear-band cavity. The second option is that the shear-plane topography can cause a momentary apparent reduction of the data in Figure 4 when the two sides of the shearing parts are moving relative to each other. In this case, the spacing between the shearing parts may become sufficiently small such that the affected part of the shear-band cavity is below the detection limit. These two options are intimately linked to each other and can therefore not be distinguished at this point. However, the impact of the Dragonfly algorithm sensitivity was tested on the XRT dataset, revealing a change of the total delaminated area of at most 3%. Thus, the uncertainty on calculating the effective cross-sectional area is marginal.

## 3.2 Calculating the true stress strain curve

Extracting the total delaminated area on the shear-plane within the sample as a function of strain allows for the calculation of the true stress-strain curve of the MG (see Section 2 for details). When doing so, the change in cross-section area is also considered. This simply takes into account the change in area due to the relative motion of the upper and lower sample part (rigid body motion), which are separated by the major and system-spanning shear band on which internal delamination occurs. The change of the load-bearing area due to internal shear-band cavities can only be calculated for each of the 49 interrupted mechanical tests individually, which is why we have chosen to linearly interpolate the delaminated area over the entirety of the nominally plastic regime. Re-loading segments, as seen in Figure 2, have been removed to produce a continuous flow curve. Figure 5 displays three definitions of the uniaxial stress-strain response, namely: a) the resulting engineering stress-strain response ($\sigma_e$ vs $\varepsilon_e$), b) the stress-strain response considering only the geometric change of the load-bearing area due to rigid body motion of the upper sample half, which we here call the 'semi-true' stress-strain curve ($\sigma_{s-t}$ vs $\varepsilon_t$), and c) the true stress-strain curve taking into account both rigid body motion and internal shear-band cavities ($\sigma_t$ vs $\varepsilon_t$). In addition to taking into account the change in actual cross-sectional loadbearing from a pure geometry perspective, the semi-true curve further uses the instantaneous length of the sample for the strain determination. Therefore, the final strain exceeds the one of the engineering stress-strain curve but is the same as for the true stress-strain curve, which in addition relies on the factual load-bearing area determined from the XRT. We emphasize that this true stress-strain calculation is based on the actual axial length of the sample and the actual load-bearing area. Since strain localization into shear bands occurs, the standard true strain equation ($\varepsilon_t = \ln(l/l_{undef})$) cannot be meaningfully used.

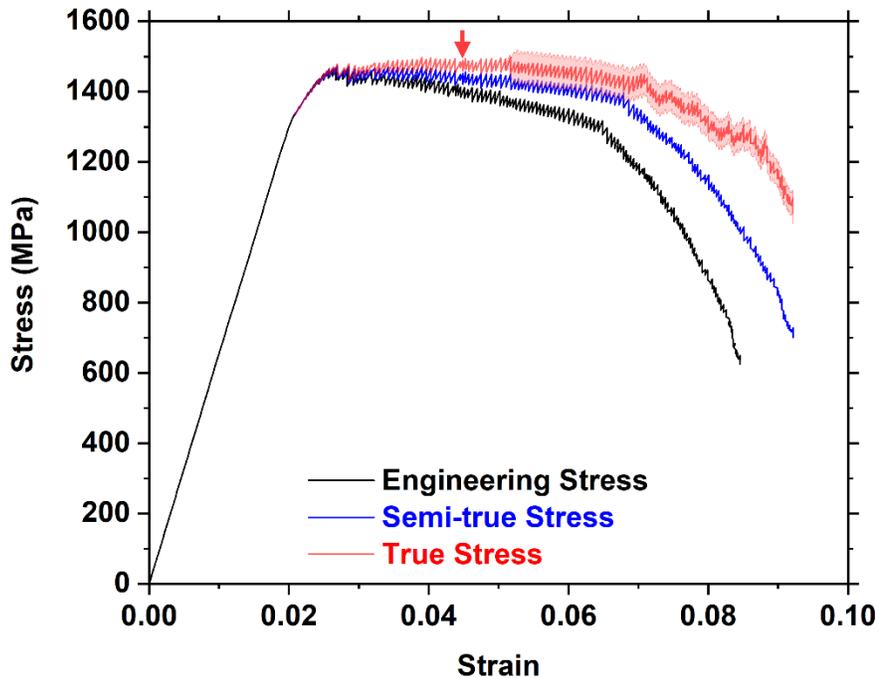

Figure 5: An engineering stress-strain curve, a 'semi-true' stress-strain curve that only takes into account geometric changes induced by each intermittent shear event, and a true stress-strain curve based on the geometric changes as well as the internal

shear-band cavity evolution. The red arrow indicates when the first resolvable change of the shear-band plane can be detected in the XRT and the red contour around the second half of the true stress-strain curve indicates the estimated error based on the XRT image reconstruction.

It can be seen that the flow stress level essentially remains constant in the true stress-strain representation at a stress level of $\sim 1450\ MPa$ until $\varepsilon_t = 0.052$ ($\varepsilon_{pl} = 0.031$). This is not the case for the engineering and the semi-true stress-strain curve, both of which exhibit a strain softening at much smaller strains, essentially starting after the extended yielding transition. The point at $\varepsilon_t = 0.068$ ($\varepsilon_{pl} = 0.047$), where the true stress begins to reduce significantly can now clearly be linked to the appearance of the first shear-band cavity (early stages of cavity 1 in Figure 3d). After a continued smooth strain-softening behavior due to a gradual increase of the total delaminated area, temporary fluctuations in the overall stress level of the serrated flow signature begin to emerge at $\varepsilon_t = 0.068$. This variation of the flow stress is due to the previously addressed non-monotonic evolution of the shear-band cavity area. In order to exemplify the stress fluctuations solely due to a varying sensitivity of the XRT segmentation algorithm, the corresponding error due to a 3% varying shear-band cavity area is indicated along the later part of the true stress-strain curve where XRT reveals the evolution of a shear-band cavity structure. At the final strain, the stresses are 627 MPa, 701 MPa, and 1056 MPa for the engineering data, the 'semi-true' data, and the true-stress strain data, respectively. Both the 'semi-true' ($\sigma_{s-t}$ vs $\varepsilon_t$) and true ($\sigma_t$ vs $\varepsilon_t$) stress-strain curve naturally exhibit larger final strains because the instantaneous sample length is considered. Relative to the yield stress, the engineering stress reduced by more than factor of two (56%), whereas the decay is only about 25% for the calculated true stress. Despite the correction for both geometry and internal shear-band cavitation, there is thus still a considerable softening present, which partly is due to possible delamination that is beyond the resolution limit, but which also is due to a true intrinsic strain-softening of the shear-band material, as the accompanying AE-data will support.

## 3.3    Assessment of possible damage mechanisms via in-situ acoustic emission

In order to combine the XRT imaging data with a probe that is sensitive to structural dissipation processes, the interrupted mechanical testing was correlated with time-synchronized AE-streaming during each loading sequence. As demonstrated for reload #18 in Figure 6, each stress-drop is accompanied with an AE-pulse that is emitted by the sample at the very onset of the stress drop. This feature has been the object of investigation in previous studies [28, 30-32] and was linked to crackling noise associated with the formation or reactivation of a shear band. From this perspective, the underlying source mechanisms of the pulse is thus of dilatational nature, and selected pulse characteristics allow quantifying the structural volume expansion of the shear-band material [33]. In other words, AE-sensing was used as a probe that gives insight into the stress-driven glass transition that the shear-band material undergoes prior to shear-band propagation.

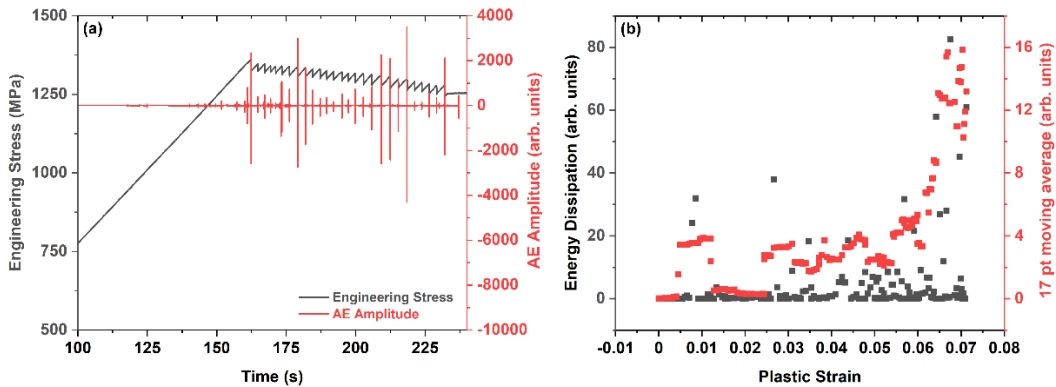

Figure 6: (a) The stress-time curve from reloading #18 accompanied by the time-synchronized AE-data revealing AE-pulses at each onset of a stress drop. (b) The dissipated energy as a function of plastic strain for the individual stress drops of all loading segments (left y-axis) and for a smoothed moving average (right y-axis) to emphasize the gradual rise with increasing strain.

Since XRT revealed internal microcracking events on the shear-band plane, it becomes possible that two different source mechanisms give rise to the recorded crackling noise. These are shear dilatation (the volume expansion of the shear-band material during shear) as well as microcracking (shear-band cavitation) during flow, the latter of which is also known to give rise to acoustic emission pulses in crystalline metals [33]. To investigate this question further, we have quantified the pulse amplitude, rise time, pulse duration, pulse energy, and the pulse frequency spectrum using continuous wavelet transforms. Relating these pulse properties to strain or any strain-dependent property, as for example the shear-band cavity volume, allows only identifying that the dissipated energy increases on-average as a function of plastic strain (Figure 6b). The fact that no other correlation can be revealed, strongly suggests that the probed metrics of the AE-pulses are not linked to the progressive internal delamination process, which therefore lack or do not emit waves within the range of the used broad-band sensor [28]. In view of the softened shear-band material during shear, it is most likely that the growth of the shear-band cavities proceeds gradually and that any microcrack growth within the softened material does neither result in recordable frequencies nor amplitudes. A similar understanding applies to the absence of AE-pulses during shear-band propagation [31], even though this earlier work did not consider the two different damage mechanisms identified here. However, Figure 6b does provide some fundamental insight into the underlying shear-band initiation process. The formation of a shear band requires bond breaking, first of the as-cast structure upon initial shear-band formation and subsequently of the relaxed shear-band material upon shear-band re-initiation. After each stress drop, shear-band aging sets in [34] that allows relaxation of the rejuvenated material within the shear band. This means, each AE-pulse stemming from shear-band initiation has an energy related to the structural damage process that reactivates flow. Figure 6b demonstrates that this energy, quantified as the pulse amplitude squared integrated over the pulse time from the beginning ($t_o$) to the peak amplitude ($t_o + t_r$), does increase with plastic strain. Integrating the rising part of the pulse can be used as a measure for the total volume change $\delta V$ of the underlying dilatational source under the condition that the sensor parameters ($C$) remain constant [30]: $\delta V = 1/C \int_{t_0}^{t_0+t_r} u(t)dt$, where $u(t)$ is the surface displacement from the measured wave. As such, the data shown in Figure 6b further indicates that the magnitude of the relative volumetric change during shear-band initiation increases with strain, providing a fundamental structural reason for the intrinsic softening of the true stress-strain curve in Figure 5.

## 4. Discussion

## 4.1. The true stress-strain curve of a metallic glass

The above summarized findings demonstrate how internal microcracking proceeds during shear-band mediated plastic flow of metallic glasses, adding an evolutionary insight to recent static tomographic observations obtained prior to failure [11, 21-23]. Numerous shear-band cavities form across the shear-band plane, of which the area increases strongly with strain. An approximate exponential growth is suggested by the data, but the limited strain range does not allow a definite determination of the precise functional growth form. The precise strain-dependent characterization of the shear-band cavities allows determining a true stress-strain curve of a metallic glass, which shows that macroscopic strain-softening is linked to the onset of cavity formation. This suggests that the large variety of post-yielding behaviour, as highlighted in Figure 1, is a manifestation of differently evolving internal shear-band cavity morphologies. Considering specifically the curves in Figure 1, the regimes of inhomogeneous plastic flow at a stable stress level vary, exhibiting onsets of strain softening between shortly after yield to about 7% engineering strain. Whilst not demonstrated for multiple samples here, the different strain-softening rates are likely a result of varying evolving shear-band cavity morphologies. It is expected that true stress-strain curves, if calculated as demonstrated here, would show a strongly reduced variation from sample to sample. Based on this, it becomes clear that the critical post-yield property of MGs is not the often-quoted low strain at failure but instead the unpredictability of the failure strain. Indeed, a low but precisely predicted plastic failure strain would be a reliable mechanical design criteria.

Strain hardening is absent for all curves displayed in Figures 1 and 5 but is often reported in the literature [18, 35, 36]. Some previous studies have proposed an intrinsic mechanism of strain hardening, which may be related to free-volume annihilation [12, 13, 35], but the more widely accepted interpretation is that shear-band interaction can result in increased shear resistance (hardening) and thus flow stress [37]. The latter geometric argument requires a high shear-band density, which naturally is linked to larger failure strains. The convolution of geometric effects (both sample shape and instrumental alignment) and intrinsic damage evolution rarely allow unambiguous conclusions, motivating the approach of single shear-band plasticity pursued here.

However, previous work highlighting the final internal shear-band cavity morphology in a number of samples [11] has revealed that true stresses can significantly increase beyond the yield stress of the material. This increase in stress was attributed to a roughness dependent friction between the delaminated surfaces. The contribution of this frictional force becomes increasingly relevant with the magnitude and roughness of the delaminated area, which amounted to to $50 - 94\%$ at $\varepsilon = 0.13 - 0.15$ in the previously reported work. For example, in the case of mild steels roughness-dependent friction leads to an increase of the frictional coefficient from 0.07 to 0.58 for an increase in center line average roughness from $0.02 \mu m$ to $0.6 \mu m$ [38]. Following this rationale, a model taking into account a friction coefficient that causes a frictional resistance to shear on the shear plane was able to explain the apparent hardening observed in these previous results for MGs. The experiments presented here, revealed only a delaminated area of ca. 34% at a true strain of 0.094 and the rms roughness, determined via the XRT data, of the shear-band cavity area is about four to six times lower than in the earlier work that exhibited apparent strain hardening based on the true stress data. Taking these differences into account, the model of apparent hardening due to frictional resistance does indeed not yield any noticeable frictional contribution to the material's shear resistance, owing to the lower roughness and smaller delaminated area. We hence conclude that in cases where the delaminated area and its roughness are low enough, frictional resistance would not play a significant role and the calculated true stress is much more a reflection of the inherent shear resistance of the flowing material inside the shear band. Besides the flow-response of the shear-band material, it is thus the large shear-plane roughness – a property that cannot be controlled - that will contribute to the precise true stress-strain behaviour. Consequently, the plastic stress-strain evolution of MGs will always be much more variable than for crystalline alloys due to the varying shear-band cavity morphology and roughness that develops during yielding.

Despite this geometric uncertainty, the in-situ AE-analysis allows an isolated insight into the structural change of the flowing material. The case of a true stress-strain curve in Fig. 5 with a continuous strain softening demonstrated here may not be entirely free from shear-band topography constraints that can affect the measured stress level. However, given the monotonic softening these effects can only be minor. As such, it becomes interesting to consider the origin of the softening, which is determined by the shear-band material itself. Based on the data in Figure 6b and its associated analysis, the structure admitting volumetric shear dilatation exhibits relative expansion magnitudes with increasing shear strain that amount to a maximum of ca. 7-10 % assuming a mean shear-band thickness in the range of 20-50 nm and are therefore approximately in line with previous work [21, 30, 39, 40]. Since the reference state is unknown, it is only possible to use the strain-dependent local yield stress at each serration as a proxy for the structural state prior to shear-band re-initiation. This stress decreases while at the same time the volume expansion during structural softening increases on average (Figure 6b). The AE-signal does hence provide a good argument for that the pronounced strain softening is due to atomic scale damage accumulation during shear. Albeit an indirect measure underlying the assumption of a dilatational point source, this supports recent findings of a strain-dependent shear-band density decrease (free volume increase) derived from scanning transmission electron microscopy [41].

### 4.2. The fractal nature of shear-band cavities

Tracking the strain-dependent evolution of the shear-band cavities offers an unprecedented insight into the evolution of the two surfaces that eventually will be revealed after fracture. With the continued formation and growth of shear-band cavities and therefore reduction of the actual load-bearing area, it is expected that most of the surface roughness of the final fracture surfaces is already determined along the plastic part of the stress-strain curve. This realization offers evaluating the geometrical parameters of the observed shear-band cavities with respect to self-similarity and possible fractal behaviour, the latter of which has been a central feature in fracture of a variety of structural materials [42, 43], including metallic glasses [44]. Central to these works has been the slit island method, first described by Mandelbrot [45], when studying fracture surfaces of steel. In this seminal work, the power-law equation $Area \propto Perimeter^n$ returned a value of $n = 1.56$. As $n = 2/d_f$, a material specific fractal dimension $d_f = 1.28$ is derived. A similar approach has been applied to fracture surfaces of MGs, where damage cavities were identified from the mismatch of two opposite fractured sample pieces, yielding a fractal dimension of $d_f = 1.43$ ($n = 1.40$) [44]. Here we exploit the strain-dependent tracking of cavity parameters and plot the cavity area as a function of cavity boundary length for the in Figure 3d highlighted five cases. A clear power-law scaling with a scaling exponent of $n = 1.68$ ($d_f = 1.19$) is observed, as shown in Figure 7. It can be seen that this power-law is robust even at the level of an individual shear-band cavity, where up to two orders of magnitude along this scaling behavior is covered by one cavity (#5). In other words, the scaling can be revealed by tracking a single shear-band cavity as a function of strain.

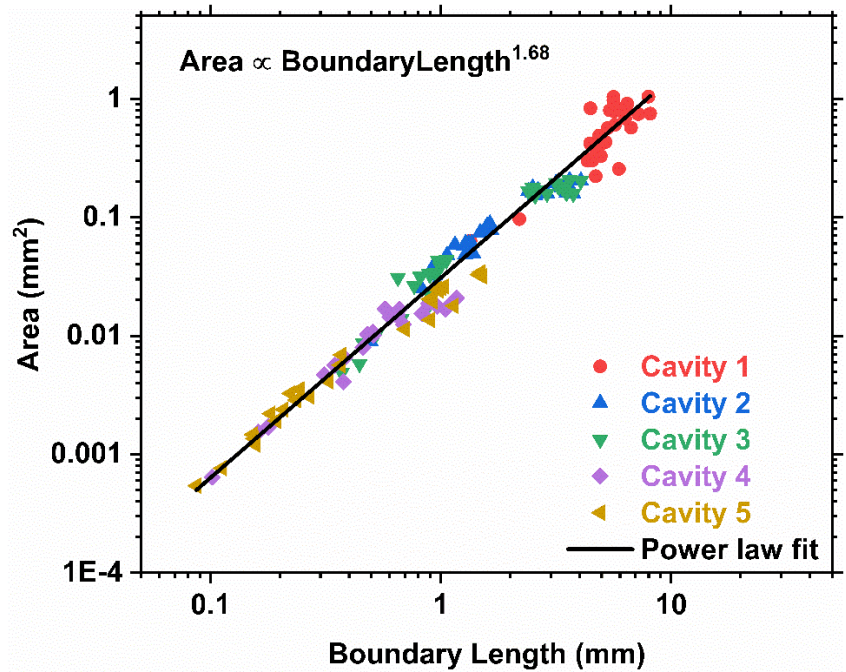

Figure 7: Shear-band cavity area as a function of shear-band cavity boundary length reveals a power-law scaling.

One must therefore conclude that the fractal nature of the data seen in Figure 7 significantly preempts the formation of the fracture surface. In the present work, no classical fracture experiment with a pre-existing crack that subsequently propagates from one sample side to the other was conducted. Instead, shear-band cavities forming shortly after yielding and that are distributed across the shear-band plane are the seeds for final failure. We emphasize that failure was not reached here. At the atomic scale, coalescence of nano-voids ahead of the crack tip are believed to facilitate crack propagation [46, 47], whereas the here tracked mesoscopic shear-band cavities were proposed to develop due to high local tensile stress components that form when the two sample halves, separated by the shear-band plane, slide relative to each other [48]. It is the large shear-band plane roughness that must give rise to such stress localization triggering cavitation. This shear-band plane roughness is set when the system spanning defect forms during yielding. Nucleating across the shear-band plane during shear, eventual percolation of cavities will mark the critical moment at which failure occurs.

One signature of critical cavity percolation is the fracture surface roughness, which has been studied in detail for many different solids [43]. In particular, nano-scale surface corrugations perpendicular to the crack-propagation direction haven been reported [49]. Being inaccessible to direct observation in a crack-propagation experiment, continuum mechanics simulations have demonstrated how cavitation can emerge from specifically introduced weak sites and that the shear-band path connecting these can give rise to the experimentally observed periodic nano-corrugations [50]. Analyzing post-failure fracture surfaces containing these corrugations and other roughness features relies on the Frasta [51] or related optical methods to reconstruct damage cavities and to quantitatively evaluate surface-roughness fluctuations. Such efforts have demonstrated self-similar fracture surface behavior with a variety of universal or non-universal roughness exponents, also referred to as the Hurst exponent $H$. Relevant to the scaling of Figure 7 is that $d_f$ of the area-boundary-length scaling can directly be related to $H$ via $H = 2 - d_f$. This relationship relates a local property ($d_f$) to a global property ($H$) of fracture, allowing us to determine $H \approx 0.8$, which is in excellent agreement with rapid unstable crack propagation in a number of solids [52, 53].

Having derived two fundamental properties known from statistical fracture surface analysis, our results raise the question of how to interpret a compressive stress-strain curve of a MG. One the one hand side, plastic flow proceeds via shear banding in all parts of the shear-plane that is not cavitating, but on the other hand the strain-dependent formation and growth of damage cavities identifies local fracture. We therefore conclude that plastic flow and fracture essentially coexist along the entirety of the plastic stress-strain curve. While cycles of shear-driven disordering and thermally-activated shear-band aging occurs in the shear band [34], formed shear-band cavities grow via crack-front propagation mechanisms in the wake of which the self-affine fracture surface forms. As such, the variability of post-yielding stress-strain behavior is due to an internal stable fracture process that may begin shortly after yielding and continues over a significant plastic strain until damage-cavity percolation initiates failure.

## 5. Conclusions

The strain-dependent evolution of internal shear-band cavities was traced with x-ray tomography and in-situ acoustic emission along the stress-strain curve. The obtained data allows calculating the true load-bearing area and therefore the true stress-strain response by taking into account internal micro-cracking that occurs on the shear-band plane. The conducted experiments reveal the following:

- Within the resolution of the experiments, strain-softening coincides with the onset of internal microcracking (formation of shear-band cavities) on the shear-band plane.
- Shear-band cavities form at various plastic strains at different locations on the shear-band plane and grow approximately exponentially.
- True stress-strain curves that are free from geometric effects soften, indicating intrinsic softening of the shear-band material.
- With continued plastic strain, the in-situ acoustic emission signal reveals an on average increasing relative volume dilatation.
- The strain-dependent shear-band cavity area and boundary length describe a power-law scaling with a fractal dimension of 1.19 that in turn yields a roughness exponent of ca. 0.8, both of which are in excellent agreement with post-failure fracture surface analysis of similar materials.
- Plastic deformation and fracture evolution are concluded to co-exist along the entire plastic stress-strain curve of a metallic glass.

The collected data explains the origin of the large variability of stress-strain behaviours reported for metallic glasses. It is the spatially distributed nucleation and growth of shear-band cavities at various plastic shear strains that strongly dictates the true load-bearing area of the tested specimens. This means, the post yielding response of metallic glasses will always be much more subject to its individual internal damage evolution than their crystalline counterparts. Generalized conclusions on post yielding parameters, such as flow stress or strain at failure, therefore demand a sufficiently large number of experiments to truly reflect heterogeneous internal microcracking.

## 6. Acknowledgement

This research was carried out in part at the Frederick Seitz Materials Research Laboratory Central Research Facilities, and the Imaging Technology Group at the Beckman Institute, University of Illinois. R.M. is grateful for financial support from the Department of Materials Science and Engineering at UIUC, and the Federal Institute for Materials Research and Testing (BAM). A.D. gratefully acknowledges financial support from the American Society for Nondestructive Testing (ASNT) and 3M during his graduate work.

## 7. Data Availability

The raw/processed data required to reproduce these findings cannot be shared at this time as the data also forms part of an ongoing study.

## 8. Author contributions

**A. Das**: Formal analysis, Investigation, Writing - Original Draft, Visualization **C. Ott**, **D. Pechimuthu and R. Moosavi**: Software, Formal analysis, Investigation **M. Stoicha**: Resources, Review & Editing

**P. M. Derlet**: Writing – Review & Editing **R. Maaß**: Writing – Original Draft, Review & Editing, Supervision, Resources, Funding acquisition